\documentclass[10pt, conference]{IEEEtran}
\usepackage[ruled]{algorithm2e}
\usepackage{algpseudocode}
\usepackage[space]{cite}
\usepackage{flushend}
\usepackage{amsfonts}
\usepackage{amsmath}
\usepackage{graphicx}
\usepackage{subfigure}
\usepackage{caption}
\usepackage{multicol}
\usepackage{multirow}

\SetAlFnt{\small}
\SetAlCapFnt{\small}
\SetAlCapNameFnt{\small}
\SetAlCapHSkip{0pt}

\begin{document}

\title{On Secure and Usable Program Obfuscation: A Survey}
\author{\IEEEauthorblockN{
Hui Xu\IEEEauthorrefmark{2},
Yangfan Zhou\IEEEauthorrefmark{3},
Yu Kang\IEEEauthorrefmark{3},
Michael R. Lyu\IEEEauthorrefmark{2}
}
\IEEEauthorblockA{
\IEEEauthorrefmark{2} Dept. of Computer Science, The Chinese University of Hong Kong}
\IEEEauthorblockA{\IEEEauthorrefmark{3} School of Computer Science, Fudan University}
}
\maketitle
\thispagestyle{plain}
\pagestyle{plain}

\begin{abstract}
\textit{Program obfuscation} is a widely employed approach for software intellectual property protection.  However, general obfuscation methods (\textit{e.g.,} lexical obfuscation, control obfuscation) implemented in mainstream obfuscation tools are heuristic and have little security guarantee.  Recently in 2013, Garg \textit{et al.} have achieved a breakthrough in secure program obfuscation with a graded encoding mechanism and they have shown that it can fulfill a compelling security property, \textit{i.e.,} indistinguishability.  Nevertheless, the mechanism incurs too much overhead for practical usage.  Besides, it focuses on obfuscating computation models (\textit{e.g.,} circuits) rather than real codes.  In this paper, we aim to explore secure and usable obfuscation approaches from the literature.  Our main finding is that currently we still have no such approaches made secure and usable.  The main reason is we do not have adequate evaluation metrics concerning both security and performance.  On one hand, existing code-oriented obfuscation approaches generally evaluate the increased obscurity rather than security guarantee.  On the other hand, the performance requirement for model-oriented obfuscation approaches is too weak to develop practical program obfuscation solutions.

\end{abstract}
\maketitle

\section{Introduction} \label{sec:intro}
\textit{Program obfuscation} is a major technique for software intellectual property protection~\cite{cappaert2012code}.  It transforms computer programs to new versions which are semantic-equivalent with the original one but harder to understand.  The concept was originally introduced at the International Obfuscated C Code Contest in 1984, which awarded creative C source codes with ``smelly styles''.  It now becomes an indispensable technique for software protection.  There are dozens of code obfuscation ideas proposed in the literature and implemented by obfuscation tools.  However, a truth we cannot ignore is that current mainstream obfuscation techniques do not provide a security guarantee.

An obfuscation approach is secure if it guarantees that the essential program semantics can be protected and demonstrates adequate hardness for adversaries to recover the semantics.  Existing obfuscation approaches generally cannot meet such criteria.  Moreover, there are many notable attacks on current obfuscation mechanisms (\textit{e.g.,}~\cite{udupa2005deobfuscation,chandrasekharan2005deobfuscation,guillot2010automatic,coogan2011deobfuscation,
yadegari2015generic,bichsel2016statistical}).  Such attacks generally assume particular obfuscation mechanisms and directly attack them without need to solve any hard problems.  Take the most recent attack by Bichsel \textit{et al.}~\cite{bichsel2016statistical} as an example, which recovers a significant portion of the original lexical information from obfuscated Android apps.  The attack just employs machine learning techniques to predict the original strings leveraging the residual information.  It seems that the security of program obfuscation is not well-established as other security primitives, such as cryptography.  So one important question is ``\textit{do we have secure and usable program obfuscation approaches? if not, what is the viable means towards one?}''

\begin{figure}[t]
\centering
\includegraphics[width=0.49\textwidth]{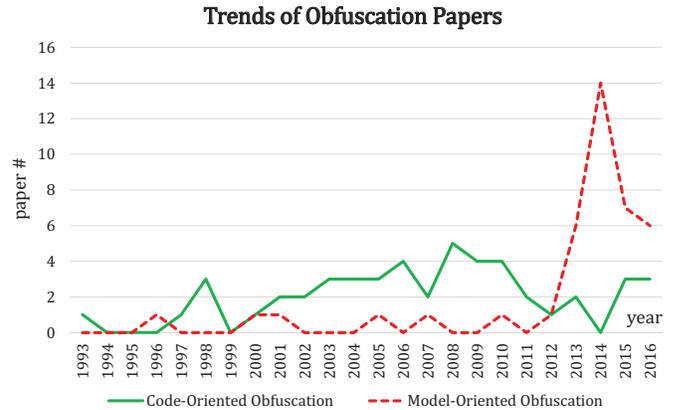}
\caption{The distribution of our surveyed obfuscation paper across years.}
\label{fig:distribution}
\end{figure}

Recently in 2013, a breakthrough came from the theoretical perspective.  Garg \textit{et al.}~\cite{garg2013candidate} proposed the first candidate program obfuscation algorithm (\textit{i.e.,} graded encoding) for all circuits and showed that it could achieve a compelling security property: \textit{indistinguishability}.  The idea is to encode circuits with multilinear maps.  It has been inspiring many follow-up investigations which aim to deliver obfuscation approaches with provable security (\textit{e.g.,}~\cite{zimmerman2015obfuscate,lewi20165gen}).  Figure~\ref{fig:distribution} demonstrates an explosion of such obfuscation research with a dashed line.  However, such graded encoding approaches are still too inefficient to be usable.  Besides, they focus on obfuscating computation models, such as circuits or Turing Machines, rather than real codes.  Although circuits and codes are closely related, graded encoding mechanisms cannot be applied to practical codes directly.  We need to figure out the gaps and connections in between, which are essential to explore secure and usable obfuscation approaches.

This survey aims to explore secure and usable program obfuscation approaches from the literature.  Firstly, we study whether usable code-oriented obfuscation approaches can be secure.  We confirm that no such approaches have demonstrated well-studied security.  The primary reason is that current evaluation metrics are not adequate for security purposes.  Existing investigations generally adopte the metrics proposed by Collberg~\textit{et~al.}~\cite{collberg1998manufacturing}, which are potency, resilience, stealthy and cost.  Note that such evaluation metrics emphasize on the increased obscurity (\textit{i.e.,} potency) rather than the semantics that remained clear in an obfuscated program.  Therefore, the metrics guarantee little security.  

Secondly, we study whether we can develop usable program obfuscation approaches from existing model-oriented obfuscation investigations.  The result is negative.  Current graded encoding mechanisms are too inefficient to be usable.  They only satisfy the performance requirement defined by Barak \textit{et al.}~\cite{barak2001possibility}, \textit{i.e.,} an obfuscated program should incur only polynomial overhead.  The requirement might be too weak because a qualified program can still grow very large.  Moreover, existing model-oriented obfuscation approaches are only applicable to real programs which contain only simple mathematical operations; they do not apply to ordinary codes with complex syntactic structures.  For example, model-oriented obfuscation approaches do not consider some code components, \textit{e.g.,} the lexical information and API calls.  Such components serve as essential clues for adversaries to analyze a program and should be obfuscated.

To summarize, this paper serves as a first attempt to explore secure and usable obfuscation approaches with a comparative study of code-oriented obfuscation and model-oriented obfuscation.  Our result is that we have no secure and usable obfuscation approaches in current practice.  To develop such approaches, we suggest the community to design appropriate evaluation metrics at first.
 
The rest of this paper is organized as follows: we first discuss the related work in Section~\ref{sec:literature}; then we introduce our study approach in Section~\ref{sec:approach} and major results in Section~\ref{sec:result}; we survey the literature about code-oriented obfuscation in Section~\ref{sec:code_obf} and model-oriented obfuscation in Section~\ref{sec:model_obf};  then we discuss the possible paths towards secure and usable program obfuscation in Section~\ref{sec:discussion};  finally, we conclude this study in Section~\ref{sec:conclusion}.

\section{Related Work}\label{sec:literature}

As obfuscation has been studied for almost two decades, several surveys are available.  However, they mainly focus on either code-oriented obfuscation or model-oriented obfuscation.  The surveys of code-oriented obfuscation include~\cite{balakrishnan2005code,majumdar2006survey,drape2009intellectual,roundy2013binary,schrittwieser2016protecting}.  Balakrishnan and Schulze~\cite{balakrishnan2005code} surveyed several major obfuscation approaches for both benign codes and malicious codes.  Majumdar~\textit{et~al.}~\cite{majumdar2006survey} conducted a short survey that summarizes the control-flow obfuscation techniques using opaque predicates and dynamic dispatcher.  Drape~\textit{et~al.}~\cite{drape2009intellectual} surveyed several obfuscation techniques via layout tranformation, control-flow tranformation, data tranformation, language dependent transformations, \textit{etc}.  Roundy~\textit{et~al.}~\cite{roundy2013binary} systematically studied obfuscation techniques for binaries, which have been frequently used by malware packers; Schrittwieser~\textit{et~al.}~\cite{schrittwieser2016protecting} surveyed the resilience of obfuscation mechanisms to reverse engineering techniques.  The surveys of model-oriented obfuscation include~\cite{horvath2016birth,barak2016hopes}.  Horvath~\textit{et~al.}~\cite{horvath2016birth} studied the history of cryptography obfuscation, with a focus on graded encoding mechanisms.  Barak~\cite{barak2016hopes} reviewed the importance of indistinguishability obfuscation.

To our best knowledge, none of the existing surveys includes a thorough comparative study of code-oriented obfuscation and model-oriented obfuscation.  Indeed, the two categories are closely related, because they frequently cite each other.  For example, the impossibility result for model-oriented obfuscation in~\cite{barak2001possibility} has been widely cited by code-oriented investigations (\textit{e.g.,}~\cite{schrittwieser2011code}).  Our survey, therefore, severs as a pilot study on synthesizing code-oriented obfuscation and model-oriented obfuscation.

Note that there are another two papers~\cite{dalla2005semantic,kuzurin2007concept} that have noticed the gaps between code-oriented obfuscation and model-oriented obfuscation, and they work towards secure and usable obfuscation.  Preda and Giacobazzi~\textit{et~al.}~\cite{dalla2005semantic} proposed to model the security properties of obfuscation with abstract interpretation, which can be further employed to deliver obfuscation solutions (\textit{e.g.,}~\cite{dalla2006opaque, dalla2007code}).  Kuzurin~\textit{et~al.}~\cite{kuzurin2007concept} noticed that current security properties for model obfuscation are too strong, and they proposed several alternative properties for practical program obfuscation scenarios.  Note that such papers coincide with us on the importance of our studied problem.  We will discuss more details in Section~\ref{sec:discussion}.

\section{Study Approach}\label{sec:approach}
\begin{figure}[t]
\centering
\includegraphics[width=0.49\textwidth]{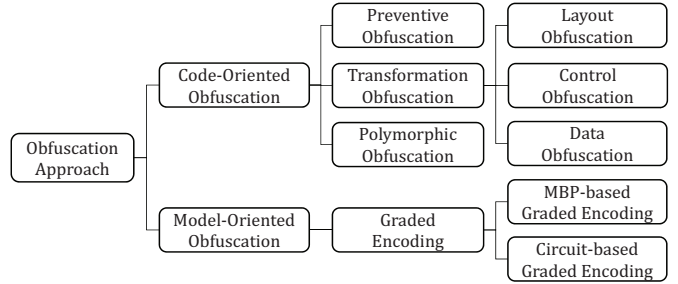}
\caption{A taxonomy of program obfuscation approaches.}
\label{fig:taxonomy}
\end{figure}

\subsection{Survey Scope}

This work discusses program obfuscation, including both code-oriented obfuscation and model-oriented obfuscation.  A program obfuscator is a compiler that to transforms a program $\mathbb{P}$ into another version $\mathcal{O}(\mathbb{P})$, which is functionally identical to $\mathbb{P}$ but much harder to understand.  Note that the concept is consistent with the definitions of code-oriented obfuscation by Collberg~\textit{et~al.}~\cite{collberg1997taxonomy} and model-oriented obfuscation by Barak~\textit{et~al.}~\cite{barak2001possibility}.  By this concept, we rule out some manual obfuscation approaches that can not be generalized or automated in compilers (\textit{e.g.,}~\cite{madou2006loco}).

Moreover, we restrict our study to general-purpose obfuscation, which means the obfuscation has no preference on program functionalities.  However, if some obfuscation approaches are not designed general programs but are valuable to obfuscate general programs (\textit{e.g.,} white-box encryption~\cite{chow2002white,chow2002white_des}, malware camouflages~\cite{borello2008code}), we would also discuss them.  Indeed, general-purpose obfuscation is a common obfuscation scenario and covers most obfuscation investigations.

Finally, we do not emphasize the differences among programming languages, such as C, or Java.  Such differences are not critical issues towards secure and usable obfuscation.  If one obfuscation approach is studied several times for different programming languages, we think such investigations are similar and only discuss a representative one.

\subsection{A Taxonomy of Obfuscation Approaches}
\subsubsection{Taxonomy Hierarchy}
For all the obfuscation approaches within the scope, we draw a taxonomy hierarchy as shown in Figure~\ref{fig:taxonomy}.  In the first level, we divide program obfuscation into \textit{code-oriented obfuscation} and \textit{model-oriented obfuscation}.  Each category can be identified with a groundbreaking paper.

The groundbreaking paper of code-oriented obfuscation was published in 1997 when Collberg~\textit{et~al.}~\cite{collberg1997taxonomy} conducted a pilot study on the taxonomy of obfuscation transformation.  They have discussed several transformation approaches and evaluation metrics.  Since then, obfuscation has been receiving extensive attentions in both research and industrial fields.  Followup investigations mainly propose new ideas on obfuscation transformations in layout level, control level, or data level.  Besides, there are also preventive obfuscation (\textit{e.g.,}~\cite{linn2003obfuscation,chan2004advanced,popov2007binary,darwish2010stealthy}) and polymorphic obfuscation (\textit{e.g.,}~\cite{cohen1993operating,lin2009polymorphing,hui2016nvo}).  We defer our discussion about the details to Section~\ref{sec:code_obf}. 

For model-oriented obfuscation, the groundbreaking paper was published in 2001 when \textit{Barak et~al.}~\cite{barak2001possibility} initiated the first theoretical study on the capability of program obfuscation.  They studied how much semantic information can be hidden at most via obfuscation.  To this end, they proposed a virtual black-box property and showed that not all programs can be obfuscated with the property.  Hence, what security properties can obfuscation guarantee and how to achieve the property are the two important problems of this area.  Currently, graded encoding~\cite{garg2013candidate} is the only available mechanism that can be implemented for obfuscation.  We defer our discussion about the details to Section~\ref{sec:model_obf}.

Note that other investigations (\textit{e.g.,}~\cite{kuzurin2007concept}) may employ practical obfuscation and theoretical obfuscation as the category names.  However, such a categorization approach may not be very discriminative because practical investigations on obfuscating codes may also include theoretical studies (\textit{e.g.,}~\cite{appel2002deobfuscation,ogiso2003software}), and \textit{vice versa}.  Therefore, our categorization approach with code-oriented obfuscation and model-oriented obfuscation should be more appropriate.

\subsubsection{The Differences between Code-Oriented Obfuscation and Model-Oriented Obfuscation}

\begin{table*}[ht]
\small
\caption{The differences between code obfuscation and model obfuscation}
\label{tab:gaps}
\centering 
\newcommand{\tabincell}[2]{\begin{tabular}{@{}#1@{}}#2\end{tabular}}
\renewcommand{\arraystretch}{1.5}
\begin{tabular}{|c|c|c|c|}
\hline
 \multicolumn{2}{|c|}{}& \textbf{Code-Oriented Obfuscation} & \textbf{Model-Oriented Obfuscation} \\
 \hline
 \multicolumn{2}{|c|}{\textbf{Research Community}} & Software security, software engineering, \textit{etc} & Theoretical computation, cryptography, \textit{etc} \\
\hline
 \multirow{4}{*}{\textbf{\tabincell{c}{Obfuscation \\ Problem}}} & Program to Obfuscate & Codes & Circuits or Turing Machines \\
\cline{2-4}
& Adversarial Purpose & Task-dependent & \tabincell{c}{Semantics that computes outputs \\ given inputs}\\
\cline{2-4}
& Protection Purpose & To increase program obscurity & To hide mathematical computations\\
\cline{2-4}
& Security-related Metrics & \tabincell{c}{\textit{Potency} or increased obscurity, \\ \textit{resilience} against automated attackers, \\ \textit{stealthy} to human attackers\\} & \tabincell{c}{Leaked information measured as the \\ probability of guessing $<$input, output$>$}\\
\hline
\multicolumn{2}{|c|}{\textbf{Security Basis}} & \tabincell{c}{Hard program analysis problems\\ (\textit{e.g.,} pointer analysis)} & \tabincell{c}{Cryptography algorithms \\(\textit{e.g.,} multilinear maps)}\\
\hline
\end{tabular}
\end{table*}

Code-oriented obfuscation demonstrates non-trivial gaps with model-oriented obfuscation as summarized in Table~\ref{tab:gaps}.  They are studied by different research communities.  Code-oriented obfuscation interests software security experts or engineers, who deal with real software protection issues.  Model-oriented obfuscation interests scientists who pursue theoretical study on circuits or Turing Machines.  Besides, model-oriented obfuscation also interests cryptographers because the candidate obfuscation approaches are based on cryptographic primitives.  

The problems of code-oriented obfuscation and model-oriented obfuscation are very differet.  Firstly, real codes are more complex than general computation models.  For example, it may include components that are not considered in circuits, such as lexical information and function calls.  Such components may serve as essential information for adversaries to interprete the software and should be obfuscated.  Besides, it may contain challenging issues for obfuscators to handle, such as concurrent operations and pointers.  Secondly, the two categories demonstrate different attacker models.  Code obfuscation assumes the adversarial purpose is to understand a released software program, and the attacking methods can be either automated deobfuscation tools or manual inspections.  Such obfuscation approaches are generally evaluated with the increased obscurity or program complexity, resilience to automated attackers, stealthy to human attackers, and costs.  On the other hand, model-oriented obfuscation only assumes automated attackers (\textit{e.g.,} probabilistic polynomial-time Turing Machine) and it assumes the adversarial purpose is to infer the functionality of a computation model (circuit or Turing Machine).  Such investigations (\textit{e.g.,}~\cite{barak2001possibility}) generally made assumptions that a program can be represented as explicit pairs of $<$input, output$>$.  In this way, one can evaluate the security with mathematical representations, \textit{i.e.,} the probability of guessing the $<$input, output$>$ pairs from an obfuscated program.  A negligible probability implies that the obfuscated program leaks little information or it is secure.

Finally, the two categories resort to different security basis when proposing secure obfuscation solutions.  Code-oriented obfuscation is interested in hard program analysis problems, such as pointer analysis.  Model-oriented obfuscation generally makes further assumptions that a model contains only basic mathematical operations.  In this way, it can adopt cryptographic primitives based on some mathematical hardness assumptions, such as multilinear maps.

Note that the differences are summarized based on existing obfuscation investigations until this survey is published.  Some gaps may be mitigated in the future.

\subsection{Secure and Usable Obfuscation} \label{sec:secure}
To facilitate the following study, we clarify the concept of secure and usable obfuscation.  An obfuscation approach is secure if it performs well in two aspects: \textit{obfuscation effectiveness} and \textit{resistance}.  Obfuscation effectiveness means the confidentiality of program semantics.  A secure obfuscation approach should hide as much program semantics as possible, especially the essential semantics.  Resistance means the hardness to recover the confidential semantics.  A secure obfuscation approach should be resistant to attacks.  The two aspects are consistent with current evaluation metrics in both code-oriented obfuscation and model-oriented obfuscation fields.  For code-oriented obfuscation, obfuscation effectiveness is similar as potency, while resistance includes both resilience and stealth.  For model-oriented obfuscation, effectiveness means the security property (\textit{e.g.,} indistinguishability) which defines the maximum information that an obfuscated program leaks; while resistance means the complexity in attacking an obfuscation algorithm.

An obfuscation is usable if it can be applied to obfuscate real codes, and the incurred performance overhead is acceptable for real application scenarios.  Such overhead includes both program size and execution time.  The performance of an obfuscation approach can be arguably acceptable if it incurs trivial overhead.

\section{Overview of Findings} \label{sec:result}
In short, we have no secure and usable program obfuscation approaches to date.  The primary reason is that we do not have adequate evaluation metrics concerning both security and usability.

Firstly, none of the existing code-oriented obfuscation investigations evaluate residual semantics in an obfuscated program.  They generally adopt the evaluation metrics proposed by Collberg~\textit{et~al.}~\cite{collberg1998manufacturing}, which judges the increased obscurity rather than the unprotected code semantics.  None of them employs evaluation metrics can meet our security requirement, especially in obfuscation effectiveness.  Therefore, designing appropriate metrics seems the priority for developing secure obfuscation approaches in the future.

Secondly, current model-oriented obfuscation approaches are too inefficient to be employed in practice.  Such approaches can satisfy the performance requirement defined by Barak et al.~\cite{barak2001possibility}, \textit{i.e.,} the obfuscated program incurs only polynomial overhead, but they still cost too much.  Such a performance requirement with polynomial overhead is too weak for an approach to be practical.  Besides, the security requirements (\textit{e.g.,} indistinguishability) for obfuscating circuits might be too rigid to develop obfuscation solutions, not to mention a practical one.  This may justify why graded encoding is the only obfuscation approach to date that can meet a security requirement.  Moreover, current model-oriented obfuscation mechanisms are only applicable to codes with simple mathematical operations.  Neither do we know how to handle non-mathematical code syntax nor how to handle other programming concepts, such as control flows and data flows.

\section{Code-Oriented Obfuscation}\label{sec:code_obf}
In 1993, Cohen~\cite{cohen1993operating} published the first code obfuscation paper.  Later in 1997, Collberg~\textit{et~al.}~\cite{collberg1997taxonomy} conducted a groundbreaking study on the taxonomy of obfuscation transformations.  Then, many obfuscation approaches were proposed in the literature.

Based on the purpose of protection, we divide code-oriented obfuscation approaches into three categories: preventive obfuscation, transformation obfuscation, and polymorphic obfuscation.  Preventive obfuscation aims to impede attackers from obtaining the real codes.  Transformation obfuscation degrades the readability of the real codes.  Polymorphic obfuscation aims to prevent attackers from locating targeted semantics or features in each obfuscated version.  Transformation obfuscation serves as a major code obfuscation technique, and most of the obfuscation approaches fall into this category.  We further divide them into layout transformation, control transformation and data transformation, each of which focuses on increasing the obscurity of a particular perspective.

\subsection{Preventive Obfuscation}
Preventive obfuscation raises the bar for adversaries to obtain code snippets in readable formats.  It is generally designed for non-scripting programming languages, such as C/C++ and Java.  For such software, a disassembly phase is required to translate machine codes (\textit{e.g.,} binaries) into human readable formats.  Preventive obfuscation, therefore, aims to obstruct the disassembly phase by introducing errors to general dissemblers.

Linn and Debray~\cite{linn2003obfuscation} conducted the first preventive obfuscation study on ELF (executable and linkage format) programs, or binaries.  They propose several mechanisms to deter popular disassembling algorithms. To thwart linear sweep algorithms, the idea is to insert uncompleted instructions as junk codes after unconditional jumps.  The mechanism takes effect if a disassembler cannot handle such uncompleted instructions.  To thwart recursive algorithms, they further replace regular procedure calls with branch functions and jump tables.  In this way, the return addresses are only determined during runtime, and they can hardly be known by static disassemblers.  Similarly, Popov~\textit{et~al.}~\cite{popov2007binary} have proposed to convert unconditional jumps to traps which raise signals.  Then they employ a signal handling mechanism to achieve the original semantics.  Darwish et~al.~\cite{darwish2010stealthy} have verified that such obfuscation approaches are effective against commercial disassembly tools, \textit{e.g.,} IDA pro~\cite{ida}.  

The idea is also applicable to the decompilation process of Java bytecodes.  Chan and Yang~\cite{chan2004advanced} proposed several lexical tricks to impede Java decompilation.  The idea is to modify bytecodes directly by employing reserved keywords to name variables and functions.  This is possible because the validation check of identifiers is only performed by the frontend.  In this way, the modified program can still run correctly, but it would cause troubles for decompilation tools.

To measure the effectiveness of preventive obfuscation, Linn and Debray~\cite{linn2003obfuscation} proposed \textit{confusion factor}, \textit{i.e.,} the ratio of incorrectly disassembled instructions (or blocks, or functions) to all instructions~\cite{linn2003obfuscation}.  Popov~\textit{et~al.}~\cite{popov2007binary} also adopted the metrics.  Besides, they proposed another factor that measures the ratio of correct edges on a control-flow graph.  Such approaches are based on tricks.  Although they may mislead existing disassembly or decompilation tools, they are vulnerable to advanced handmade attacks.

\subsection{Layout Obfuscation} 
Layout obfuscation scrambles a program layout while keeping the syntax intact.  For example, it may change the orders of instructions or scramble the identifiers of variables and classes.

Lexical obfuscation is a widely employed layout obfuscation approach which transforms the meaningful identifiers to meaningless ones.  For most programming languages, adopting meaningful and uniform naming rules (\textit{e.g.,} Hungarian Notation~\cite{simonyi1999hungarian}) is required as a good programming practice.  Although such names are specified in source codes, some would remain in the released software.  For example, the names of global variables and functions in C/C++ are kept in binaries, and all names of Java are reserved in bytecodes.  Because such meaningful names can facilitate adversarial program analysis, we should scramble them.  To make the obfuscated identifiers more confusing, Chan~\textit{et~al.}~\cite{chan2004advanced} proposed to deliberately employ the same names for objects of different types or within different domains.  Such approaches have been adopted by ProGuard~\cite{proguard} as a default obfuscation scheme for Android programs.

Besides, several investigations study obfuscation via shuffling program items.  For example, Low~\cite{low1998protecting} proposed to seperate the related items of Java programs wherever possible, because a program is harder to read if the related information is not physically close.  Wroblewski~\cite{wroblewski2002general} proposed to reorder a sequence of instructions if it does not change the program semantics.

In general, layout obfuscation has promising resistance because some transformations are one-way which cannot be reversed. But the obfuscation effectiveness is only limited to layout level.  Moreover, some layout information can hardly be changed, such as the method identifiers from Java SDK.  Such residual information is essential for adversaries to recover the obfuscated information.  For example, Bichsel~\textit{et~al.}~\cite{bichsel2016statistical} tried to deobfuscated ProGuard-obfuscated apps, and they successfully recovered around 80\% names.

\subsection{Control Obfuscation}
Control obfuscation increases the obscurity of control flows.  It can be achieved via introducing bogus control flows, employing dispatcher-based controls, and \textit{etc}.

\subsubsection{Bogus Control Flows}
Bogus control flows refer to the control flows that are deliberately added to a program but will never be executed.  It can increase the complexity of a program, \textit{e.g.,} in McCabe complexity~\cite{mccabe1976complexity} or Harrison metrics~\cite{harrison1981complexity}.  For example, McCabe complexity~\cite{mccabe1976complexity} is calculated as the number of edges on a control-flow graph minus the number of nodes, and then plus two times of the connected components.  To increase the McCabe complexity, we can either introduce new edges or add both new edges and nodes to a connected component.

To guarantee the unreachability of bogus control flows, Collberg~\textit{et~al.}~\cite{collberg1997taxonomy} proposed the idea of opaque predicates.  They defined opaque predict as the predicate whose outcome is known during obfuscation time but is difficult to deduce by static program analysis.  In general, an opaque predicate can be constantly true ($P^T$), constantly false ($P^F$), or context-dependent ($P^?$).  There are three methods to create opaque predicates: numerical schemes, programming schemes, and contextual schemes.

\vspace{0.2cm}
\textit{Numerical Schemes}
\vspace{0.1cm}

\begin{figure*}[thb]
\centering
\subfigure[Opaque constant.]{
\label{fig:opq_constant}
\includegraphics[width=0.14\textwidth]{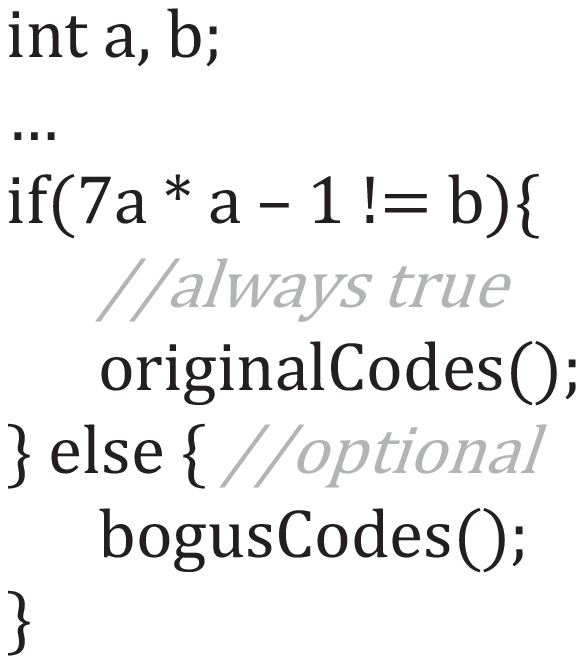}}
\hspace{0.5cm}
\subfigure[Collatz conjecture.]{
\label{fig:opq_collatz}
\includegraphics[width=0.232\textwidth]{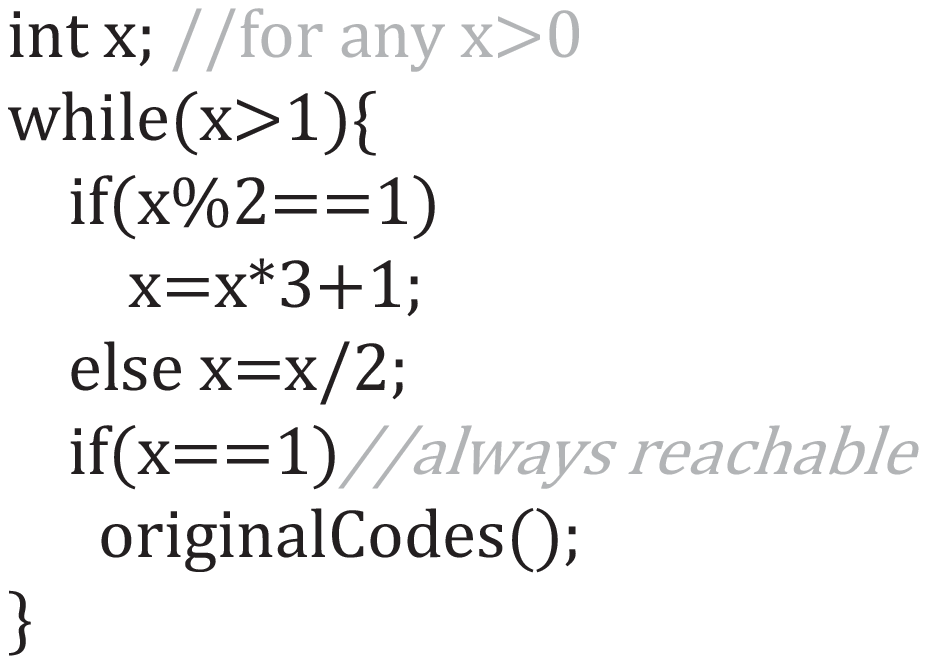}}
\hspace{0.5cm}
\subfigure[Dynamic opaque predicate.]{
\label{fig:opq_dynamic}
\includegraphics[width=0.332\textwidth]{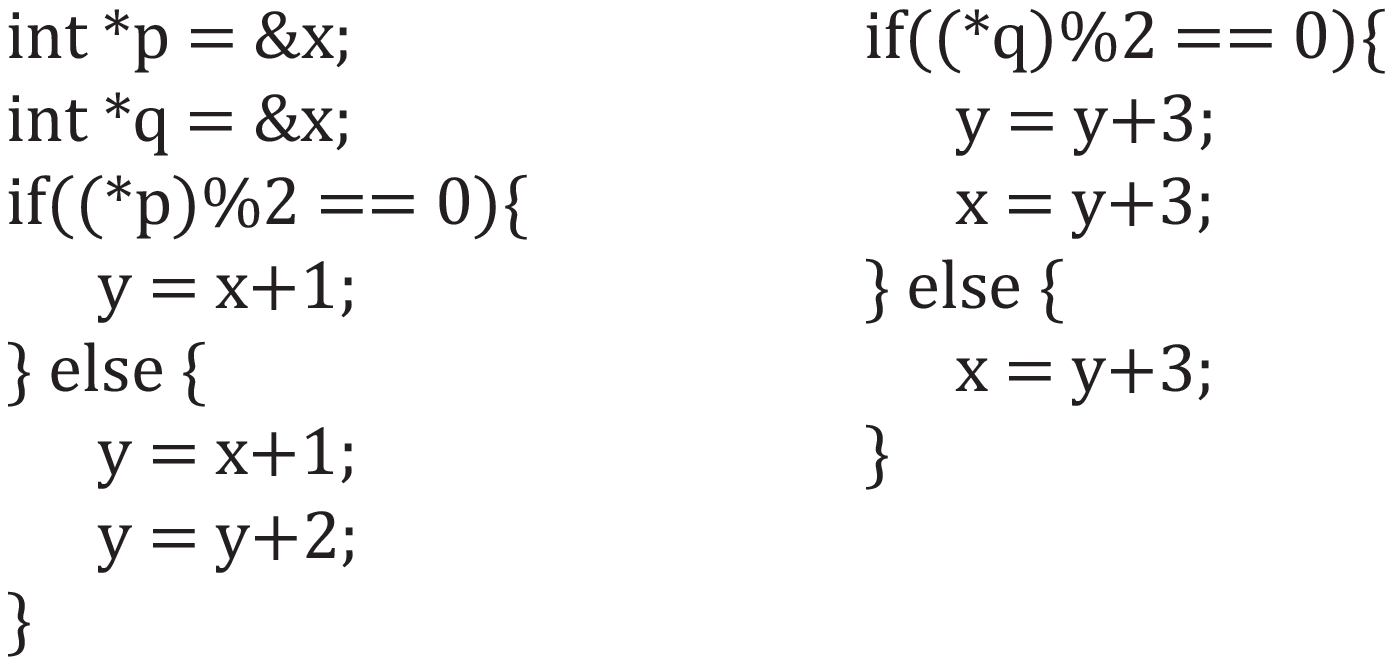}}
\caption{Control obfuscation with opaque predicates.}
\label{fig:opq_numerical}
\end{figure*}

Numerical schemes compose opaque predicates with mathematical expressions.  For example, $7x^2-1 \neq y^2$ is constantly true for all integers $x$ and $y$.  We can directly employ such opaque predicates to introduce bogus control flows.  Figure~\ref{fig:opq_constant} demonstrates an example, in which the opaque predicate guarantees that the bogus control flow (\textit{i.e.,} the else branch) will not be executed. However, attackers would have higher chances to detect them if we employ the same opaque predicates frequently in an obfuscated program.  Arboit~\cite{arboit2002method}, therefore, proposed to automatically generate a \textit{family} of such opaque predicates, such that an obfuscator can choose a unique opaque predicates each time.

Another mathematical approach with higher security is to employ \emph{crypto functions}, such as hash function $\mathcal{H}$~\cite{sharif2008impeding}, and homomorphic encryption~\cite{zhu2005provable}.  For example, we can substitute a predicate $x==c$ with $\mathcal{H}(x)==c_{hash}$ to hide the solution of $x$ for this equation.  Note that such an approach is generally employed by malware to evade dynamic program analysis.  We may also employ crypto functions to encrypt equations which cannot be satisfied.  However, such opaque predicates incur much overhead.

To compose opaque constants resistant to static analysis, Moser~\textit{et~al.}~\cite{moser2007limits} suggested employing 3-SAT problems, which are NP-hard.  This is possible because one can have efficient algorithms to compose such hard problems~\cite{selman1996generating}.  For example, Tiella and Ceccato~\cite{tiella2017automatic} demonstrated how to compose such opaque predicates with k-clique problems.

To compose opaque constants resistant to dynamic analysis, Wang~\textit{et~al.}~\cite{wang2011linear} propose to compose opaque predicates with a form of \emph{unsolved conjectures} which loop for a number of times.  Because loop is a challenging issue for dynamic analysis, the approach in nature should be resistant to dynamic analysis.  Examples of such conjectures include Collatz conjecture, $5x+1$ conjecture, Matthews conjecture.  Figure~\ref{fig:opq_collatz} demonstrates how to employ Collatz conjecture to introduce bogus control flows.  No matter how we initialize $x$, the program terminates with $x=1$, and \texttt{originalCodes()} can always be executed.

\begin{figure*}[t]
\centering
\subfigure[Souce code.]{
\label{fig:wang_flat_a}
\includegraphics[width=0.13\textwidth]{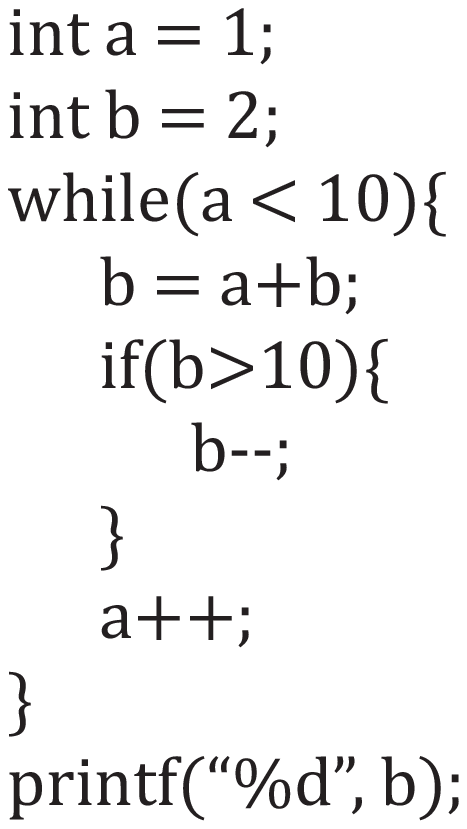}}
\hspace{0.8cm}
\subfigure[Dismantling while.]{
\label{fig:wang_flat_b}
\includegraphics[width=0.15\textwidth]{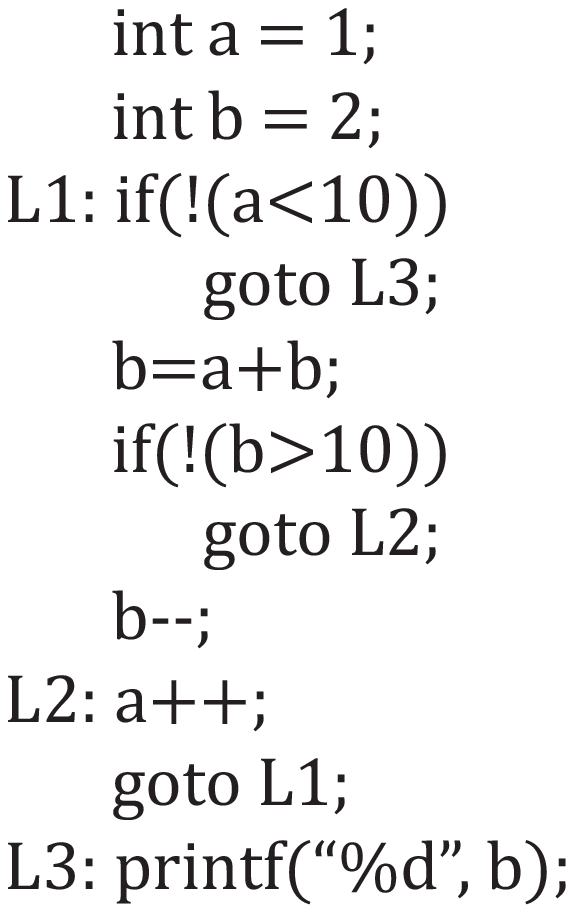}}
\subfigure[Using switch.]{
\label{fig:wang_flat_c}
\includegraphics[width=0.50\textwidth]{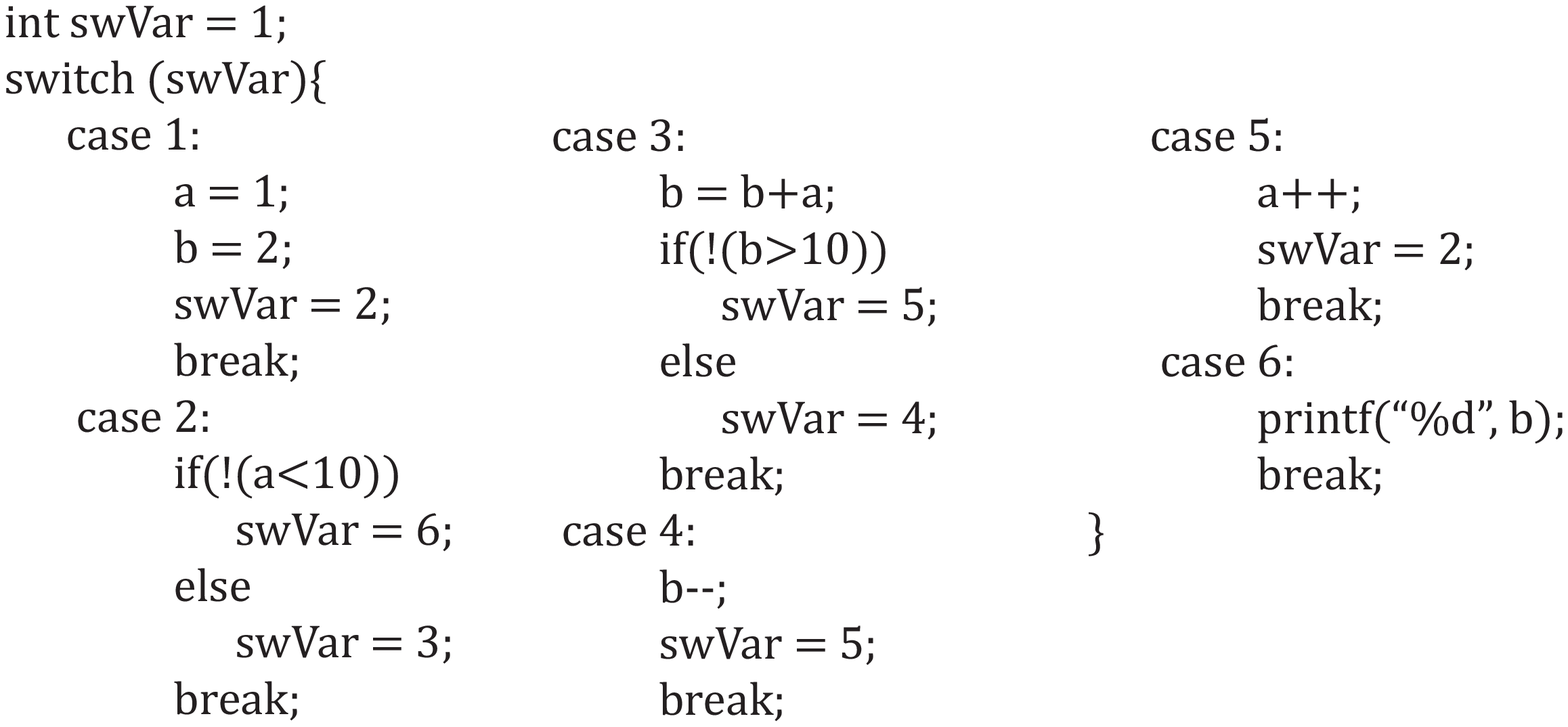}}
\caption{Control-flow flattening approach proposed by Wang~\textit{et~al.}~\cite{wang2000software}.}
\label{fig:code_flattenning}
\end{figure*}

\vspace{0.2cm}
\textit{Programming Schemes}
\vspace{0.1cm}
 
Because adversarial program analysis is a major threat to opaque predicates, we can employ challenging program analysis problems to compose opaque predicates.  Collberg~\textit{et~al.} suggest two classic problems, \textit{pointer analysis} and \textit{concurrent programs}.

In general, pointer analysis refers to determining whether two pointers can or may point to the same address.  Some pointer analysis problems can be NP-hard for static analysis or even undecidable~\cite{landi1991pointer}.  Another advantage is that pointer operations are very efficient during execution.  Therefore, one can compose resillient and efficient opaque predicts with well-designed pointer analysis problems, such as maintaining pointers to some objects with dynamic data structures~\cite{collberg1998manufacturing}.

Concurrent programs or parallel programs is another challenging issue.  In general, a parallel region of $n$ statements has $n!$ different ways of execution.  The execution is not only determined by the program, but also by the runtime status of a host computer.  Collberg~\textit{et~al.}~\cite{collberg1998manufacturing} proposed to employ concurrent programs to enhance the pointer-based approach by concurrently updating the pointers.  Majumdar~\textit{et~al.}~\cite{majumdar2006manufacturing} proposed to employ distributed parallel programs to compose opaque predicates. 

Besides, some approaches compose opaque predicates with programming tricks, such as leveraging \emph{exception handling mechanisms}.  For example, Dolz and Parra~\cite{dolz2008using} proposed to use the \texttt{try-catch} mechanism to compose opaque predicates for \texttt{.Net} and Java.  The exception events include division by zero, null pointer, index out of range, or even particular hardware exceptions~\cite{chen2009control}.  The original program semantics can be achieved via tailored exception handling schemes.  However, such opaque predicates have no security basis, and they are vulnerable to advanced handmade attacks.

\vspace{0.2cm}
\textit{Contextual Schemes}
\vspace{0.1cm}

Contextual schemes can be employed to compose variant opaque predicates(\textit{i.e.,} $\{P^?\}$).  The predicates should hold some deterministic properties such that they can be employed to obfuscate programs.  For example, Drape~\cite{drape2009intellectual} proposed to compose such opaque predicates which are invariant under a contextual constraint, \textit{e.g.,} the opaque predicate $x \text{ mod } 3 == 1$ is constantly true if $x \text{ mod } 3:1$ $?$ $x\text{++ : }x=x+3$.  Palsberg~\textit{et~al.}~\cite{palsberg2000experience} proposed dynamic opaque predicates, which include a sequence of correlated predicates.  The evaluation result of each predicate may vary in each run.  However, as long as the predicates are correlated, the program behavior is deterministic.  Figure~\ref{fig:opq_dynamic} demonstrates an example of dynamic opaque predicates.  No matter how we initialize $*p$ and $*q$, the program is equivalent to $y = x + 3, x = y + 3$.

The resistance of bogus control flows largely depends on the security of opaque predicates.  An ideal security property for opaque predicates is that they require worst-case exponential time to break but only polynomial time to construct.  Notethat some opaque predicates are designed with such security concerns but may be implemented with flaws.  For example, the 3-SAT problems proposed by Ogiso~\textit{et~al.}~\cite{ogiso2003software} are based on trivial problem settings which can be easily simplified.  If such opaque predicates are implemented properly, they would be promising to be secure.

\subsubsection{Dispatcher-Based Controls}
A dispatcher-based control determines the next blocks of codes to be executed during runtime.  Such controls are essential for control obfuscation, because they can hide the original control flows against static program analysis.  

One major dispatcher-based obfuscation approach is control flattening, which transforms codes of depth into shallow ones with more complexity.  Wang~\textit{et~al.}~\cite{wang2000software} firstly proposed the approach.  Figure~\ref{fig:code_flattenning} demonstrates an example from their paper that transforms a \texttt{while} loop into another form with \texttt{switch-case}.  To realize such transformation, the first step is to transform the code the into an equivalent reprensentation with \texttt{if-then-goto} statements as shown in Figure~\ref{fig:wang_flat_b}; then they modify the \texttt{goto} statements with \texttt{switch-case} statements as shown in Figure~\ref{fig:wang_flat_c}.  In this way, the original program semantics is realized implicitly by controlling the data flow of the switch variable.  Because the execution order of code blocks are determined by the variable dynamically, one cannot know the control flows without executing the program.  Cappaert and Preneel~\cite{cappaert2010general} formalized control flattening as employing a dispatcher node (\textit{e.g.,} \texttt{switch}) that controls the next code block to be executed; after executing a block, control is transferred back to the dispatcher node.  Besides, there are several enhancements for code flattening.  For example, to enhance the resistance to static program analysis on the switch variable, Wang~\textit{et~al.}~\cite{wang2001protection} proposed to introduce pointer analysis problems.  To further complicate the program, Chow~\textit{et~al.}~\cite{chow2001approach} proposed to add bogus code blocks.

L{\'a}szl{\'o} and Kiss~\cite{laszlo2009obfuscating} proposed a control flattening mechanism to handle specific C++ syntax, such as \texttt{try-catch}, \texttt{while-do}, \texttt{continue}.  The mechanism is based on abstract syntax tree and employs a fixed pattern of layout.  For each block of code to obfuscate, it constructs a while statement in the outer loop and a \texttt{switch-case} compound inside the loop.  The \texttt{switch-case} compound implements the original program semantics, and the switch variable is also employed to terminate the outer loop.  Cappaert and Preneel~\cite{cappaert2010general} found that the mechanisms might be vulnerable to local analysis, \emph{i.e.,} the switch variable is directly assigned such that adversaries can infer the next block to execute by only looking into a current block.  They proposed a strengthened approach with several tricks, such as employing reference assignment (\textit{e.g.,} $swVar = swVar + 1$) instead of direct assignment (\textit{e.g.,} $swVar = 3$), replacing the assignment via \texttt{if-else} with a uniform assignment expression, and employing one-way functions in calculating the successor of a basic block.

Besides control flattening, there are several other dispather-based obfuscation investigations (\textit{e.g.,}~\cite{linn2003obfuscation,ge2005control,zhang2010theory,schrittwieser2011code}).  Linn and Debray~\cite{linn2003obfuscation} proposed to obfuscate binaries with branch functions that guide the execution based on the stack information.  Similarly, Zhang~\textit{et~al.}~\cite{zhang2010theory} proposed to employ branch functions to obfuscate object-oriented programs, which defines a unified method invocation style with an object pool.  To enhance the security of such mechanisms, Ge~\textit{et~al.}~\cite{ge2005control} proposed to hide the control information in another standalone process and employ inter-process communications.  Schrittwieser and Katzenbeisser~\cite{schrittwieser2011code} proposed to employ diversified code blocks which implement the same semantics.

Dispatcher-based obfuscation is resistance against static analysis because it hides the control-flow graph of a software program.  However, it is vulnerable to dynamic program analysis or hybrid approaches.  For example, Udupa\textit{ et~al.}~\cite{udupa2005deobfuscation} proposed a hybrid approach to reveal the hidden control flows with both static analysis and dynamic analysis.

\subsubsection{Misc}
There are several other control obfuscation approaches that do not belong to the discussed categories.  Examples of such approaches are instructional control hiding (\textit{e.g.,} \cite{balachandran2011software,domas15}) and API call hiding (\textit{e.g.,} \cite{kovacheva2013efficient,Bohannon17}).  They generally have special obfuscation purposes or based on particular tricks.

\vspace{0.2cm}
\textit{Instructional Control Hiding}
\vspace{0.1cm}

Instructional control hiding converts explicit control instructions to implicit ones.  Balachandran and Emmanuel~\cite{balachandran2011software} found that control instructions (\textit{e.g.,} \texttt{jmp}) are important information for reverse analysis.  They proposed to substitute such instructions with a combination of \texttt{mov} and other instructions which implements the same control semantics.  In an extreme case, Domas~\cite{domas15} think all high-level instructions should be obfuscated.  He proposed \textit{movobfuscation}, which employs only one instruction (\textit{i.e.,} \texttt{mov}) to compile the program.  The idea is feasible because \texttt{mov} is Turing complete~\cite{dolan2013mov}.  

\vspace{0.2cm}
\textit{API Call Hiding}
\vspace{0.1cm}

Collberg~\textit{et~al.}~\cite{collberg1997taxonomy} proposed a problem that the function invocation codes in Java programs are well understood but hard to obfuscate.  They suggest substituting common patterns of function invocation with less obvious ones, such as those discussed by Wills~\cite{wills1990automated}.  The problem is significant, but surprisingly it has not been studied a lot by other investigations except \cite{kovacheva2013efficient,Bohannon17}.  Kovacheva~\cite{kovacheva2013efficient} investigated the problem for Android apps.  He proposed to obfuscate the native calls (\textit{e.g.,} to \texttt{libc} libraries) via a proxy, which is an obfuscated class that wraps the native functions.  Bohannon and Holmes~\cite{Bohannon17} investigated a similar problem for Windows powershell scripts.  To obfuscate an invocation command to Windows objects, they proposed to create a nonsense string first and then leverage Windows string operators to transform the string to a valid command during runtime.

\vspace{0.2cm}
\textit{More Tricks}
\vspace{0.1cm}

Collberg~\textit{et~al.}~\cite{collberg1997taxonomy} proposed several other obfuscation tricks, such as aggregating irrelevant method into one method, scattering a method into several methods.  Such tricks are also discussed in other investigations (\textit{e.g.,}~\cite{low1998protecting,madou2006effectiveness}) and implemented in obfuscation tools (\textit{e.g.,} JHide~\cite{ertaul2004jhide}).  Besides, Wang~\textit{et~al.}~\cite{wang2016translingual} proposed \textit{translingual obfuscation}, which introduces obscurity by translating the programs written in C into ProLog before compilation.  Because ProLog adopts a different program paradigm and execution model from C, the generated binaries should become harder to understand.   Majumdar~\textit{et~al.}~\cite{majumdar2007slicing} proposed slicing obfuscation, which increases the resistance of obfuscated programs against slicing-based deobfuscation attacks, such as by enlarging the size of a slice with bogus codes.

Not that all existing control obfuscation approaches focus on syntactic-level transformation, while the semantic-level protection has rarely been discussed.  Although they may demonstrate different strengths of resistance to attacks, their obfuscation effectiveness concerning semantic protection remains unclear.

\subsection{Data Obfuscation}

Data obfuscation transforms data objects into obscure representations.  We can transform the data of basic types via splitting, merging, procedurization, encoding, \textit{etc}.

\emph{Data splitting} distributes the information of one variable into several new variables.  For example, a boolean variable can be split into two boolean variables, and performing logical operations on them can get the original value.

\emph{Data merging} aggregates several variables into one variable.  Collberg~\textit{et~al.}~\cite{collberg1998breaking} demonstrated an example that merges two 32-bit integers into one 64-bit integer.  Ertaul and Venkatesh~\cite{ertaul2005novel} proposed another method that packs several variables into one space with discrete logarithms.

\emph{Data procedurization} substitutes static data with procedure calls.  Collberg~\textit{et~al.}~\cite{collberg1998breaking} proposed to substitute strings with a function which can produce all strings by specifying paticular parameter values.  Drape~\cite{drape2004obfuscation} proposed to encode numerical data with two inverse functions $f$ and $g$.  To assign a value $v$ to a variable $i$, we assign it to an injected variable $j$ as $j=f(v)$.  To use $i$, we invoke $g(j)$ instead.

\emph{Data encoding} encodes data with mathematical functions or ciphers.  Ertaul and Venkatesh~\cite{ertaul2005novel} proposed to encode strings with Affine ciphers (\textit{e.g.,} Caser cipher) and employ discrete logarithms to pack words.  Fukushima~\textit{et~al.}~\cite{fukushima2008analysis} proposed to encode the clear numbers with \texttt{exclusive or} operations and then decrypt the computation result before output.  Kovacheva~\cite{kovacheva2013efficient} proposed to encrypt strings with the RC4 cipher and then decrypt them during runtime. 

The data obfuscation ideas can also be extended to abstract data types, such as arrays and classes.  Collberg~\textit{et~al.}~\cite{collberg1998breaking} discussed the obfuscation transformations for arrays, such as splitting one array into several subarrays, merging several arrays into one array, folding an array to increase its dimension, or flattening an array to reduce the dimension.  Ertaul and Venkatesh~\cite{ertaul2005novel} suggested transforming the array indices with composite functions.  Zhu~\textit{et~al.}~\cite{zhu2006applications,zhu2007concepts} proposed to employ homomorphic encryption to obfuscate array.
Obfuscation classes is similar as obfuscation arrays.  Collberg~\textit{et~al.}~\cite{collberg1998breaking} proposed to increase the depth of the class inheritance tree by splitting classes or inserting bogus classes.  Sosonkin~\textit{et~al.}~\cite{sosonkin2003obfuscation} also discussed class obfuscation techniques via coalescing and splitting.  Such approaches can increase the complexity of a class, \textit{e.g.,} measured in CK metrics~\cite{chidamber1994metrics}.  

\subsection{Polymorphic Obfuscation}
Polymorphism is a technique widely employed by malware camouflage, which creates different copies of malware to evade anti-virus detection~\cite{xin2010misleading,you2010malware}.  It can also be employed to obfuscate programs.  Note that previous obfuscation approaches focus on introducing obscurities to one program, while polymorphic obfuscation generats multiple obfuscated versions of a program simultaneously.  Ideally, it would pose similar difficulties for adversaries to understand the components of each particular version.  It is a technique orthogonal to the classic obfuscation and mainly designed to impede large-scale and reproductive attacks to homogeneous software~\cite{forrest1997building}.

Polymorphic obfuscation generally relies on some randomization mechanisms to introduce variance during obfuscation.  Lin~\textit{et~al.}~\cite{lin2009polymorphing} proposed to generate different data structure layout during each compilation.  The data objects, such as structures, classes, and stack variables declared in functions, can be reordered randomly in each version.  Xin~\textit{et~al.}~\cite{xin2010misleading} further improved the data structure polymorphism approach by automatically discovering the data objects that can be randomized and eliminating the semantic errors generated during reordering.  Crane~\textit{et~al.}~\cite{crane2015s} proposed to randomize the tables of pointers such that the introduced diversity can be resistant to some code reuse attacks.  Besides, Xu~\textit{et~al.}~\cite{hui2016nvo} suggested introducing security features in the polymorphic code regions.

\section{Model-Oriented Obfuscation}\label{sec:model_obf}
Model-oriented obfuscation studies the theoretical obfuscation problems on computation models, such as circuits and Turing Machines.  In 1996, Goldreich and Ostrovsky~\cite{goldreich1996software} firstly studied a theoretical software intellectual property protection mechanism based on Oblivious RAM.  Later, Hada~\cite{hada2000zero} firstly studied the theoretical obfuscation problem based on Turing Machines.  In 2001, Barak~\textit{et~al.}~\cite{barak2001possibility} proposed a well-recognized modeling approach for obfuscating circuits and Turing Machines, which lays a foundation of this field.  There are two important research topics in this field: 1) what is the best security property that obfuscation can achieve? 2) how can we achieve the property?
\begin{table*}[ht]
\small
\caption{A comparison of the security properties for model-oriented obfuscation.  Notations: $\mathcal{S}$ is a polynomial-size simulator; $\mathcal{A}$ is a polynomial-size adversary; $\mathcal{L}$ is a polynomial-size learner; $\mathcal{S}_u$ is a unbounded-size simulator; $\mathbb{P}_1$ and $\mathbb{P}_2$ are programs that compute a same function and have similar cost; $\mathbb{P}$ and $\mathbb{Q}$ are programs that compute different functions; $\mathcal{S}_u^{\mathbb{P}[q(n)]}$ means querying the oracle access $\mathcal{S}_u$ for $n$ times; $\varepsilon$ is a negligible number.}
\label{tab:sec_prop}
\centering 
\newcommand{\tabincell}[2]{\begin{tabular}{@{}#1@{}}#2\end{tabular}}
\renewcommand{\arraystretch}{1.6}
\begin{tabular}{|c|c|c|}
\hline
 \textbf{Security Property} & \textbf{Requirement} & \textbf{Security Strength} \\
\hline
 \tabincell{c}{Virtual Black-Box Property (VBBP)} & \tabincell{c}{$|Pr[\mathcal{A}(\mathcal{O}(\mathbb{P})) = 1] - Pr[\mathcal{S}^\mathbb{P} = 1]|\le \varepsilon$}  & Ideal Security \\
\hline
 \tabincell{c}{Indistinguishability Property (INDP)}  & \tabincell{c}{$|Pr[\mathcal{A}(\mathcal{O}(\mathbb{P}_1)) = 1] - Pr[\mathcal{A}(\mathcal{O}(\mathbb{P}_2)) = 1]|\le \varepsilon$} & INDP $<$ VBBP \\
\hline
 \tabincell{c}{Differing-Input Property (DIP)} & \tabincell{c}{If $|Pr[\mathcal{A}(\mathcal{O}(\mathbb{P})) = 1] - Pr[\mathcal{A}(\mathcal{O}(\mathbb{Q})) = 1]| \ge \varepsilon$,\\ then $|Pr[\mathcal{A}(\mathbb{P}') = 1] - Pr[\mathcal{A}(\mathbb{Q}') = 1]| \ge \varepsilon$} & VBBP $>$ DIP $>$ INDP \\
\hline
 \tabincell{c}{Best-Possible Property (BPP)}  & \tabincell{c}{$|Pr[\mathcal{L}(\mathcal{O}(\mathbb{P}_1)) = 1] - Pr[\mathcal{S}(\mathbb{P}_2)]|\le \varepsilon$} & BPP $=$ Efficient INDP \\
\hline
 \tabincell{c}{Virtual Grey-Box Property (VGBP)} & \tabincell{c}{$|Pr[\mathcal{A}(\mathcal{O}(\mathbb{P})) = 1] - Pr[\mathcal{S}_u^{\mathbb{P}[q(n)]} = 1]| \le \varepsilon$} & INDP $<$ VGBP $<$ VBBP \\
\hline
\end{tabular}
\end{table*}
\subsection{Security Properties}
Barak~\textit{et~al.}~\cite{barak2001possibility} defined an obfuscator $\mathcal{O}$ as a ``compiler'' that inputs a program $\mathbb{P}$ and outputs a new program $\mathcal{O}(\mathbb{P})$.  $\mathcal{O}(\mathbb{P})$ should posses the same functionality as $\mathbb{P}$, incur only \textit{polynomial slowdown}, and hold some properties of unintelligibility or \textit{security}.  Note that following investigations generally adopt ``polynomial slowdown'' to discriminate whether an algorithm is efficient, but they may adopt different properties of security.  Table~\ref{tab:sec_prop} lists several major security properties discussed in the literature.  Next we introduce each property and justify why the indistinguishability property is mostly interested by existing obfuscation algorithms.

\subsubsection{Virtual Black-Box Property (VBBP)}
Ideally, an obfuscated program should leak no more information than accessing the program in a black-box manner.  The property is firstly proposed by Barak~\textit{et~al.}~\cite{barak2001possibility} as \textit{virtual black-box obfuscation}.  Let $\mathcal{A}$ be a polynomial-size adversary (\textit{e.g.,} a probabilistic polynomial-time Turing Machine).  VBBP requires that for any such adversaries, there exists a polynomial-size simulator $S$, such that $|Pr[\mathcal{A}(\mathcal{O}(\mathbb{P})) = 1] - Pr[\mathcal{S}^\mathbb{P} = 1]|$ is negligible.  The expression means any program semantics learned by the adversary can be simulated with a polynomial-size simulator.

Barak~\textit{et~al.} have shown a negative result that at least one family of efficient programs $\mathbb{P}_f(x)$ cannot be obfuscated with VBBP.  $\mathbb{P}_f(x)$ can be constructed with any one-way functions, whose semantics cannot be learned from oracle access.  Therefore, given only oracle access to the function $f(x)$, no efficient algorithm can compute $f(x)$ better than random guessing.  However, given any efficient program $\mathbb{P}'_f(x)$, there exists an efficient algorithm that can compute the function.  In this way, an efficient program that computes a property of the function can be constructed as $D(\mathbb{P}'_f(x)) \to \{0,1\}$, but it cannot be efficiently constructed from oracle access.  Goldwasser~\textit{et~al.}~\cite{goldwasser2005impossibility} and Bitansky~\textit{et~al.}~\cite{bitansky2014impossibility} further showed that some encryption programs cannot be obfuscated with VBBP when auxiliary inputs are available. 

The negative result implies we cannot achieve genreal-purpose obfuscation with VBBP.  However, it does not mean no program can be obfuscated with VBBP.  Point function is such an exception~\cite{canetti1997towards}.

\subsubsection{Indistinguishability Property (INDP) and Best-Possible Property (BPP)} 
Although VBBP is not universally attainable, we still need some attainable properties.  As an alternative, Barak~\textit{et~al.}~\cite{barak2012possibility} proposed a weaker notion: \emph{indistinguishability obfuscation}.  It requires that if two programs $\mathbb{P}_1$ and $\mathbb{P}_2$ are functionally equivalent, and they have similar program size and execution time, then $|Pr[\mathcal{A}(\mathcal{O}(\mathbb{P}_1)) = 1] - Pr[\mathcal{A}(\mathcal{O}(\mathbb{P}_2)) = 1]|$ is negligible.  Because the notion does not assume a polynomial-size simulator, it avoids the inefficiency issue caused by unobfuscatable programs.  Barak~\textit{et~al.} have shown that INDP is attainable for universal programs, such as employing an obfuscator that converts all programs to their canonical forms or lookup tables.  However, such an obfuscator might be trivial if it is inefficient.

Another important issue is how useful INDP is since it has no intuitive to hide information.  INDP guarantees that the obfuscated program leaks no more information than any other obfuscated program versions of similar cost.  Therefore, if we can design an obfuscator with INDP, it guarantees the obfuscation would have the best effectiveness.  To rule out inefficient obfuscation, Goldwasser~\textit{et~al.}~\cite{goldwasser2007best} proposed \emph{best-possible obfuscation}.  BPP requires that for any polynomial-size learner $\mathcal{L}$, there exists a polynomial-size simulator $S$, such that $|Pr[\mathcal{L}(\mathcal{O}(\mathbb{P}_1)) = 1] - Pr[\mathcal{S}(\mathbb{P}_2)]|$ is negligible.  By assuming a polynomial-size simulator, BPP excludes inefficient indistinguishability obfuscation.  Note that BPP is also similar to VBBP but it is weaker than VBBP.  The differece is that for BPP, the simulator $\mathcal{S}(\mathbb{P}_2)$ can access another version of the program, while for VBBP the simulator $\mathcal{S}^\mathbb{P}$ works as a black-box.  Lin~\textit{et~al.}~\cite{lin2016indistinguishability} proposed another similar notion \textit{exponentially-efficient indistinguishability property}(XINDP), that requires the obfuscated program should be smaller than its truth table.  Lin~\textit{et~al.} showed that XINDP implies efficient INDP under the assumption of learning with errors (LWE)~\cite{regev2005lattices}.

\subsubsection{More Properties}
There are other alternatives discussed in the literature, such as \textit{virtual grey-box property} (VGBP)~\cite{bitansky2010strong}, \textit{differing-input property} (DIP)~\cite{barak2012possibility} and its variations.  Their security levels are considered as between VBBP and INDP~\cite{horvath2016birth}.

DIP is another notion proposed by Barak \textit{et al.}~\cite{barak2012possibility}.  For two programs $\mathbb{P}$ and $\mathbb{Q}$ of the same cost, it requires if an adversary can distinguish their obfuscated versions (\textit{i.e.,} $O(\mathbb{P})$ and $O(\mathbb{Q})$), she should be able to differ any versions of $\mathbb{P}$ and $\mathbb{Q}$ with the same cost, \textit{i.e.,} to find an input $x$ such that $\mathbb{P}'(x) \neq \mathbb{Q}'(x)$.  When $\mathbb{P}$ and $\mathbb{Q}$ compute the same function, DIP implies INDP.  DIP is also known as extractability obfuscation~\cite{boyle2014extractability}.  However, Boyle and Pass~\cite{boyle2014limits}, and Garg~\textit{et~al.}~\cite{garg2014implausibility} showed that DIP is not attainable for all programs.  To avoid the impossibility, Ishai~\textit{et~al.}~\cite{ishai2015public} proposed public-coin DIP.  Note that DIP is a stronger notion than INDP, and we can have many useful applications with a DIP obfuscator~\cite{ananth2013differing}.

VGBP~\cite{bitansky2010strong} is similar to VBBP except that it empowers the simulator to unbounded size.  To be nontrival, it restricts the simulator to have only limited times of oracle access.  Bitansky \textit{et~al.}~\cite{bitansky2014virtual} shown that VGBP also implies INDP.

In brief, we cannot obfuscate universal programs with VBBP security, but we may obfuscate universal programs with INDP security.  INDP is also the best-possible security property if the obfuscation is efficient.  Therefore, if VBBP is attainable for some programs, efficient INDP would guarantee VBBP.

\subsection{Candidate Approach}
In 2013, Garg~\textit{et~al.}~\cite{garg2013candidate} proposed the first candidate algorithm to achieve INDP.  Their approach is based on the idea of \textit{functional encryption}~\cite{boneh2011functional}.  Functional encryption allows users to compute a function from encrypted data with some keys.  In the scenario of program obfuscation, suppose a program $\mathbb{P}$ computes a function $f(x)$, then the functional encryption problem is to encrypt the program $Enc(\mathbb{P})$, such that $Enc(\mathbb{P})$ can still compute $f(x)$ with a public key $K_s$, but $K_s$ should not reveal $\mathbb{P}$.  Because cryptography algorithms generally have strong security basis, functional encryption becomes a dominant idea for program obfuscation with provable security.

Currently, the only known functional encryption approach for program obfuscation is \textit{graded encoding}.  It encodes programs with multilinear maps, such that any tampering attacks that do not respect its mathematical structure (encoded with private keys) would lead to meaningless program executions.  To employ graded encoding, Garg~\textit{et~al.}~\cite{garg2013_full} proposed to convert programs to matrix branching programs (MBP) before encoding.  However, such conversion incurs much overhead.  Later, Zimmerman~\cite{zimmerman2015obfuscate}, and Applebaum and Brakerski~\cite{applebaum2015obfuscating} proposed to encode circuits directly without converting to MBPs.  Next we discuss the two mechanisms.


\subsubsection{MBP-based Graded Encoding}
An essential requirement to employ graded encoding is that the encoded programs can be evaluated.  MBP is such a model that holds a good algebraic structure for evaluation even after being encrypted.  There are two phases for MBP-based graded encoding: the first phase converts programs to MBPs; the second phase encrypts MBPs with graded encoding mechanisms.  Garg~\textit{et~al.}~\cite{garg2013_full} showed that the MBP-based approach is feasible for shallow $NC^1$ circuits and can be extended to all circuits with fully homomorphic encryption.  

\vspace{0.2cm}
\textbf{Converting to MBP}
\vspace{0.1cm}

\begin{figure}
\centering
\subfigure[A branching program.]{
\label{fig:bp}
\includegraphics[width=0.49\textwidth]{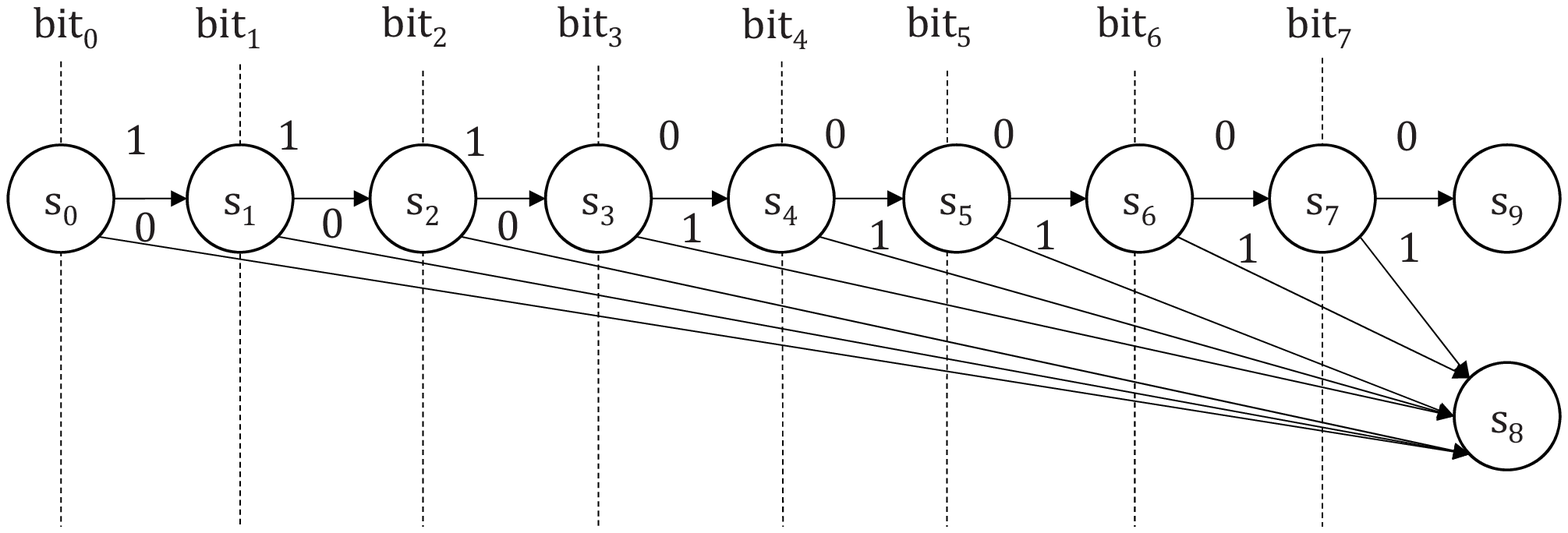}}
\subfigure[A matrix branching program.]{
\label{fig:mbp}
\includegraphics[width=0.49\textwidth]{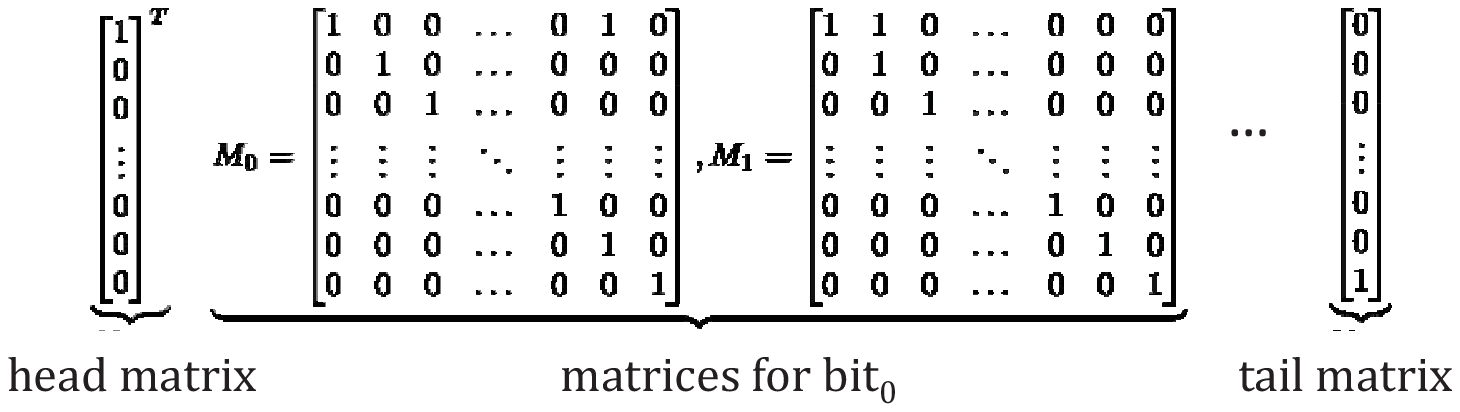}}
\subfigure[Generate randomized matrix branching program.]{
\label{fig:rmbp}
\includegraphics[width=0.49\textwidth]{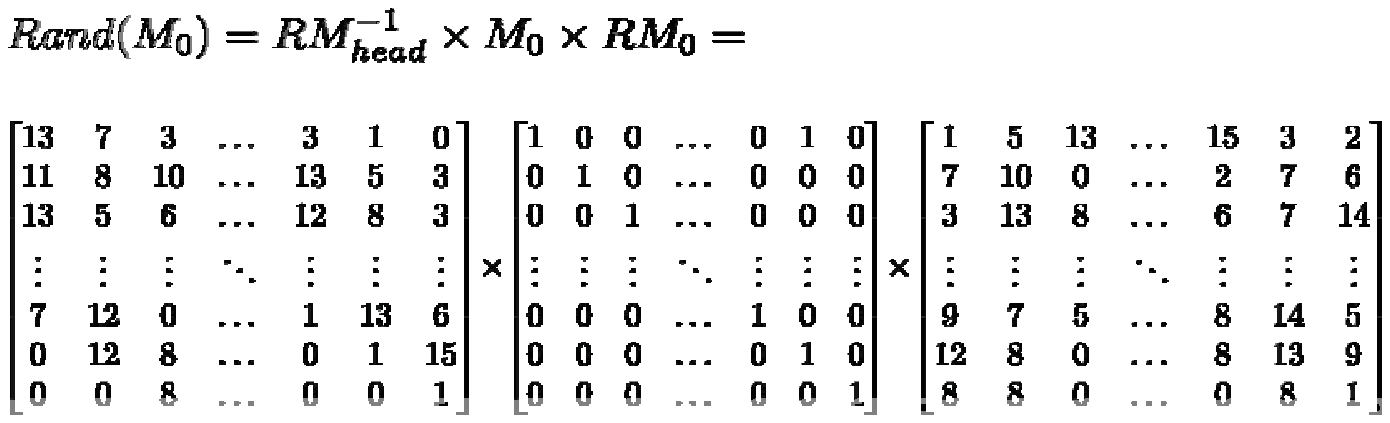}}
\caption{The procedures to convert a program (\textit{i.e.,} if $x$ of \texttt{int8} equals to $7$) to a randomized matrix branching program.}
\label{fig:enhance_rmbp}
\end{figure}

A matrix branching program that computes a function $f$ is given by a tuple
$$MBP_f=(Input, M_{head}, (M_{i,0}, M_{i,1})_{i \in l}, M_{tail})$$

$Input$ selects a matrix $M_{i,0}$ or $M_{i,1}$ for each $i$ according to the corresponding bit of input; $M_{head}$ is a row vector of size $w$; $(M_{i,0}, M_{i,1})_{i \in l}$ are matrix pairs of size $w \times w$ that encode program semantics; $M_{tail}$ is a column vector of size $w$. 

Given an input $x$, the MBP computes an output $MBP_f(x) \in \{0, 1\}$ as follows:

$$MBP_f(x) = M_{head} \times (\prod^l_{i=1}M_{i,x_{input(k)}}) \times M_{tail}$$ 

Suppose the $i$-th matrix pair corresponds to the $k$-th bit of the input.  If the $k$-th bit is 0, then $M_{i,0}$ is selected, or \textit{vice versa}.  The program output is the matrix multiplication result, which is a $1 \times 1$ matrix, or a value. 

How can we convert general programs to MBPs?  The Barrington's Theorem states that we can convert any boolean formula (boolean circuit of fan-in-two, depth $d$) to a branching program of width $5$ and length $\le 4^d$~\cite{barrington1986bounded}.  Garg~\textit{et~al.}~\cite{garg2013_full} simply employ the result and assume the MBP is composed of with $5 \times 5$ matrices.  Ananth~\textit{et~al.}~\cite{ananth2014optimizing} found the resulting MBP following Barrington's Theorem is not very efficient.  They propose a new approach that converts boolean formulas of size $s$ to matrix branching programs of width $\le 2s+2$ and length $\le s$.  Besides, there are several other efforts towards converting to more efficient MBPs, such as~\cite{sahai2014obfuscating,boneh2015semantically}.  The conversion generally includes two steps: from a program $P_f$ to a branching program $BP_f$ and from $BP_f$ to $MBP_f$.

\textit{$P_f \to BP_f$}: A branching program is a finite state machine.  For boolean formulas $P_f \in \{0,1\}$, the finite state machine has one start state, two stop states ($true$ and $false$), and several intermediate states.  Sauerhoff~\textit{et~al.}~\cite{sauerhoff1999formula} demonstrated a general approach to simulate any boolean formulas over AND and OR gates with branching programs.  It can be extended to any formulas as they can be converted to the form with only AND and OR.  Figure~\ref{fig:bp} demonstrates an example which converts a boolean program $i==7$ to a branching program.  Suppose $i$ is an integer of eight bits, the boolean formula is $b_0 \wedge b_1 \wedge b_2 \wedge \neg b_3 \wedge \neg b_4 \wedge \neg b_5 \wedge \neg b_6 \wedge \neg b_7$.  To model the branching program we need 10 states: $8$ states ($s_0$-$s_7$) that accept each bit of input, and $2$ stop states ($s_8$ for $false$, and $s_9$ for $true$).

\textit{$BP_f \to MBP_f$}:  To compose a $MBP_f$ that is functionally equivalent to $BP_f$, we should compute each matrix of the $MBP_f$.  In general, $M_{head}$ can be an all-zero row vector except the first position is $1$, and $M_{tail}$ can be an all-zero column vector except the last position is $1$.  $(M_{i,0}, M_{i,1})_{i \in len}$ can be constructed from the adjacency matrices of each state.  For example, if the first bit of input is $0$, the station transfers from $s_0$ to $s_8$, then we start with an identity matrix and assign 1 to the element of the first row and the ninth column.  Figure~\ref{fig:mbp} demonstrates the matrices corresponding to the fist input bit of Figure~\ref{fig:bp}.  

Following such converting approaches, the elements of resulting matrices are either 1 or 0.  To protect the matrices, Kilian~\cite{kilian1988founding} proposed that we can randomize the elements of an MBP while not changing its functionality.

\textit{$MBP_f \to RMBP_f$}:  To randomize the matrices, we first generate $n+1$ random integer matrices $RM_i$ and their inverse $RM_i^{-1}$ of size $w \times w$.  Then we multiply the original matrices with such random matrices as follows.

$$RM_{head}= M_{head} \times RM_{0} $$
$$RM_{0,0}= RM_{0}^{-1} \times M_{0,0} \times RM_{1} $$
$$RM_{0,1}= RM_{0}^{-1} \times M_{0,1} \times RM_{1} $$
$$...$$
$$RM_{tail}= RM_{n}^{-1} \times M_{tail}$$

The randomization mechanism ensures that all randomization matrices $RM_i$ would be canceled when evaluating $RMBP_f(x)$.  Note that to avoid errors incurred by floating-point numbers, we should guarantee all the elements of matrices are integers as shown in Figure~\ref{fig:rmbp}.  This is feasible because when the dominant of $RM_i$ is $1$, $RM_i^{-1}$ is also an integer matrix.  Stating from an identity matrix, such $RM_i$ can be obtained via iterative transformations leveraging the determinant invariant rule.

\vspace{0.2cm}
\textbf{Graded Encoding}
\vspace{0.1cm}

Garg~\textit{et~al.}~\cite{garg2013candidate} noticed that although the randomized matrix branching program provides some security, it still suffers three kinds of attacks: partial evaluation, mixed input, and other attacks that do not respect the algebraic structure.  Partial evaluation means we can evaluate whether partial programs generate the same result for different inputs.  Mixed input means we can tamper the program intentionally by selecting $M_{i,0}$ and $M_{j,1}$ if $i$ and $j$ are related to the same bit of input.  Graded encoding is designed to defeat such attacks. It is based on multilinear maps, which can be traced back to the historical multiparty key exchange problem proposed in 1976~\cite{diffie1976multiuser, boneh2003applications}.  

In general, a graded encoding scheme includes four components: \textit{setup} that generates the public and private parameters for a system, \textit{encoding} that defines how to encrypt a message with the private parameters, \textit{operations} that declare the supported calculations with encrypted messages, and a \textit{zero-testing function} that evaluates if the plain text of an encrypted message should be $0$.  Currently, there are two graded encoding schemes: GGH scheme~\cite{garg2013candidate_lattices} which encodes data over lattices, and CLT scheme~\cite{coron2013practical} which encodes data over integers.  Note that the graded encoding schemes for program obfuscation are slightly different from their original versions for multiparty key exchange.  For simplicity, below we only discuss the graded encoding schemes for obfuscation.

The \textit{GGH scheme} is named after Garg, Gentry, and Halevi~\cite{garg2013candidate_lattices}, and it is the first plausible solution to compose multilinear maps.  GGH scheme is based on ideal lattices.  It encodes an element $e$ over a quotient ring $R/\mathcal{I}$ as $e + \mathcal{I}$, where $\mathcal{I} = \langle g \rangle \subset R$ is the principal ideal generated by a short vector $g$.  The four components of GGH are defined as follows.

\textit{Setup}: Suppose the miltilinear level is $\kappa$.  The system generates an ideal-generator $g$ which is chosen as $g$ and $g^{-1}$ should be short, a large enough modulus $q$, denominators $\{z_i\}$ from the ring $R_q$.  Then we publish the zero-testing parameter as $p_{zt} = [h \prod_{i=1}^\kappa z_i/g]_q$, where $h$ is a small ring element.

\textit{Encoding}: The encoding of an element $e$ in set $S_{z_i}$ can be computed as : $u :=[(e + \mathcal{I}) / z_i]_q$.

\textit{Operations}: If two encodings are in the same set (\textit{e.g.,} $u_1:=[c_1 / z_i]_q$ and $u_2:=[c_2 / z_i]_q$), then one can add up them $u_1+u_2$.  If the two encodings are from disjoint sets, one can multiply the two encodings $u_1 \cdot u_2$.

\textit{Zero-Testing Function}: A zero testing function for a level-$\kappa$ encoding $u$ is defined as
\[ IsZero(u) =
  \begin{cases}
    1 & \quad \text{if } || [u \cdot p_{zt}]_q||_\infty \le q ^{3/4}\\
    0 & \quad \text{otherwise}\\
  \end{cases}
\]

Note that $u \cdot p_{zt} = h \cdot c /g$.  If $u$ is an encoding of $0$, $c$ should be a short vector in $\mathcal{I}$ and the product can be smaller than a threshold, otherwise, $c$ should be a short vector in some coset of $\mathcal{I}$ and the product should be very large.

The CLT scheme is another multilinear map construction approach proposed by Coron, Lepoint, and Tibouchi~\cite{coron2013practical,coron2015new}.  It is based on integers.  The four components of the scheme are defined as follows.

\textit{Setup}: The scheme generates $\kappa$ secret large primes $\{p_i\}$, small primes $\{g_i\}$, random integers $\{h_i\}$, random integers $\{z_i\}$, a modulo $q=\prod_{i=1}^{\kappa}p_i$, and a zero-testing parameter $$p_{zt} = \sum_{i=1}^{\kappa} h_i \times \prod_{i=1}^{\kappa} z_i \times g^{-1} \text{mod } p_i \times \prod_{i \neq i'} p_{i'} \text{mod } q$$

\textit{Encoding}: Suppose $r_i$ is a small random integer, the encoding of an element $e$ in set $S_{z_i}$ is $u = \frac{r_i \cdot g_i + e}{z_i}(\text{mod } p_i)$.

\textit{Operation}: If $u_i$ and $u_j$ are encodings in the same set, one can add them up. If they are from disjoint sets, they can be multiplied.

\textit{Zero-Testing Function}: A zero testing function for a level-$\kappa$ encoding $u$ is

\[ IsZero(u) =
  \begin{cases}
    1 & \quad \text{if } ||u \cdot p_{zt} (\text{mod }q)||_\infty \le q \cdot 2^{-v}\\
    0 & \quad \text{otherwise}\\
  \end{cases}
\]

In this function, $v$ is a value related to the bit-size of the encoding parameters~\cite{coron2013practical}.

Note that both the GGH scheme and CLT scheme are noisy multilinear maps, because the encoding of a value varies at different times.  The only deterministic function is the zero-testing function.  However, when a program becomes complex, the noise may overwhelm the signal.  Take the CLT scheme as an example, the size of $p_i$ should be as large as possible to overwhelm the noise.  This requirement largely restrict the usability of graded encoding.

\subsubsection{Circuit-based Graded Encoding}
Converting circuits to MBPs incurs much overhead because the size of an MBP is generally exponential to the depth of a circuit.  To avoid such overhead, Zimmerman~\cite{zimmerman2015obfuscate} and Applebaum and Brakerski~\cite{applebaum2015obfuscating} proposed to obfuscate circuit programs directly.  The approach focuses on keyed circuit families $(C(\cdot, k))_{k\in\{0,1\}^m}$, and it can be extended to general circuits because all circuits can be transformed to keyed circuits~\cite{zimmerman2014obfuscate}.  

\begin{figure}
\centering
\includegraphics[width=0.48\textwidth]{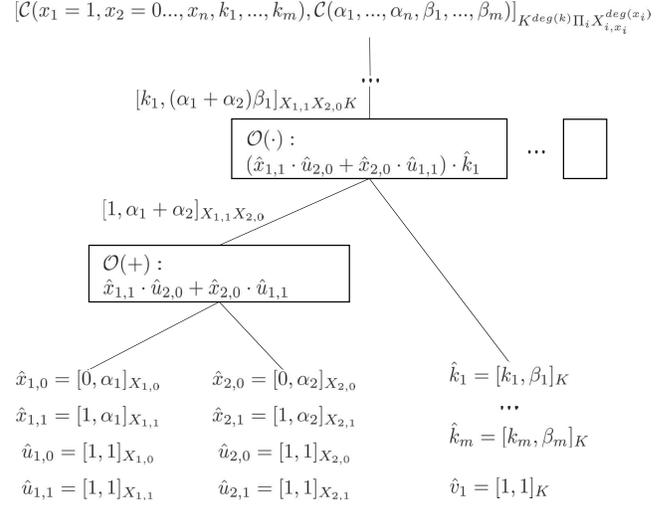}
\caption{The evaluation process of an encoded keyed circuit (key: $k_1...k_m$) given an input is $x_1=1,x_2=0...x_n$.}
\label{fig:circuit}
\end{figure}

Circuit-based graded encoding assumes a circuit structure can be made public, and only the key needs to be protected.  Figure~\ref{fig:circuit} demonstrates an example that evaluates an obfuscated circuit given an input $x_1=1,x_2=0...x_n$.  To generate such an obfuscated circuit, we encode each input wire with a pair of encodings corresponding to the input values of $0$ and $1$, \textit{i.e.,} $[x_{i,0}, \alpha_i]_{S_{i,0}}$,$[x_{i,1},\alpha_i]_{S_{i,1}}$.  To support the addition of values from different multilinear sets, we publish the encoding of $1$ for each set, \textit{i.e.,} $[1,1]_{S_{i,0}}$,$[1,1]_{S_{i,1}}$.  Besides, we encode each key bit as $[k_{i}, \beta_i]_{S_{K}}$ and publish the encoding of $1$, \textit{i.e.,} $[1,1]_{S_{K}}$.  Note that the approach introduces a checksum mechanism, \textit{i.e.,} $x_i \text{ mod } N \equiv x_i' \text{ mod } N_{eval} \equiv \alpha_i \text{ mod } N_{chk}$, s.t. $N = N_{eval} \cdot N_{chk}$.  Evaluating the obfuscated circuit just follows the original circuit structure.  As a result, we can compute an evaluation value $C(x_1,...,x_n,k_1,...,k_m) \in S_{N_{eval}}$ and a checksum $C(\alpha_1,...,\alpha_n,\beta_1,...,\beta_m) \in S_{N_{chk}}$.

\section{On Secure and Usable Obfuscation}\label{sec:discussion}
In this section, we carefully justify the prospects of the existing investigations towards secure and usable obfuscation.

\subsection{With Code-Oriented Obfuscation}
In general, existing code-oriented obfuscation approaches are usable but insecure.  We may hope to achieve secure program obfuscation in the future if we have adequate evaluation metrics for security.  Our primary supporting evidence is two folds.  Firstly, current security metrics are inadequate.  Secondly, code obfuscation techniques are promising to be resistant because deobfuscation also suffers limitations.

\subsubsection{Inadequate Security Metrics}
To our best knowledge, all existing evaluation metrics emloyed by code-oriented obfuscation investigations are inadequate concerning the security discussed in Section~\ref{sec:secure}.   Most investigations and tools (\textit{e.g.,} Obfuscator-LLVM~\cite{junod2015obfuscator}) adopt the metrics defined by Collberg~\textit{et~al.}~\cite{collberg1998manufacturing}.  Besides, other evaluation metrics (\textit{e.g.,}~\cite{anckaert2007program,ceccato2008towards,ceccato2009effectiveness,ceccato2015large}) proposed in the literature are also inadequate.  For example, Anckaert~\textit{et~al.}~\cite{anckaert2007program} followed the idea of potency and developed more detailed measurements.  Ceccato~\textit{et~al.}~\cite{ceccato2008towards,ceccato2009effectiveness} proposed to conduct controlled code comprehension experiments against human attackers, such that we can measure the security with task completion rate and time.  Such metrics are either heuristic or consider little about protecting essential program semantics.

\subsubsection{Limitations of Deobfuscation}
The most recognized deobfuscation attacks (\textit{e.g.,}~\cite{raber2007deobfuscator,guillot2010automatic}) are based on program analysis and pattern recognition, both of which suffer limitations.  Pattern recognition requires a predefined pattern repository, and it cannot automatically adapt to new obfuscation techniques.  Program analysis suffers many challenges.  For example, symbolic execution is one major program analysis approach to detect opaque predicates~\cite{yadegari2015symbolic,yadegari2015generic,ming2015loop,banescu2017predicting}, but it is vulnerable to many challenges, such as handling symbolic arrays and concurrent programs~\cite{xu2017concolic}.   

Moreover, The Rice's Theorem~\cite{hopcroft2006automata} implies that automated attackers would suffer theoretical limitations because whether a deobfuscated program is equivalent to the obfuscated version is undecidable.  Only when the tricks are known, some deobfuscation problems can be in NP~\cite{appel2002deobfuscation}.  Therefore, program obfuscation approaches are promising to have good resistance.

\subsection{With Model-Oriented Obfuscation}
Current model-oriented obfuscation approaches can be considered as secure but unusable.  We may hope to achieve some usable obfuscation applications if the performance metrics can be improved or the security requirement can be weakened.  Our evidence is two folds.  Firstly, there are several obfuscation implementations and applications which demonstrate the usability issue.  Secondly, existing investigations are optimistic about the security of model-oriented obfuscation.

\subsubsection{Usability Issues}
As the development of model-oriented obfuscation, several investigations begin focusing on application issues, such as \cite{apon2014implementing,lewi20165gen,carmer20175gen,halevi2017implementing}.  Apon~\textit{et~al.}~\cite{apon2014implementing} implemented a full obfuscation solution based on the CLT graded encoding.  It can obfuscate programs written in SAGE~\cite{stein2005sage}, a Python library for algebraic operations.  Due to the performance issue, they only demonstrated single-bit \texttt{identity} gate, \texttt{AND} gate, \texttt{XOR} gate, and point functions.  Even for an 8-bit point function, it takes hours to obfuscate the program and several minutes to evaluate the program.  Besides, the size of the resulting program is several gigabytes.  Lewi~\textit{et~al.}~\cite{lewi20165gen} implemented another obfuscator that can run on top of either libCLT~\cite{coron2013practical} or GGHLite~\cite{langlois2014gghlite,albrecht2014implementing}, which are open-source libraries of multilinear maps.  The input program should be written in Cryptol~\cite{lewis2003cryptol}, which is a programming language for design cryptography algorithms.  They also evaluated the performance when obfuscating point functions.  With a better hardware configuration, it can obfuscate 40-bit point functions in minutes and evaluate the program in seconds.  However, the obfuscated program sizes are hundreds of megabytes or even several gigabytes.  Their results show that CLT has better performance over GGH for small-size point functions, but the advantage declines when the program size grows.  Halevi~\textit{et~al.}~\cite{halevi2017implementing} implemented a simplified version of the graph-induced multi-linear maps~\cite{gentry2015graph} which should outperform the CLT scheme when the number of branching program states grows.  However, their evaluation results have not shown fundamental changes of the performance.

To summerize, we can find two usability issues in such investigations.  Firstly, the costs are unacceptable even when obfuscating straightforward mathematical expressions.  The other issue is that current obfuscation implementations only focus on elementary mathematical expressions, such as \texttt{XOR}, point functions, and conjecture normal forms~\cite{brakerski2013obfuscating,brakerski2014black}.  We do not know how to handle other advanced mathematical operations, not to mention complex code syntax.

\subsubsection{Security of Graded Encoding}
Graded encoding is very powerful, Sahai and Waters~\cite{sahai2014use} showed that INDP obfuscation can serve as a center for many cryptographic applications.  Moreover, several investigations (\textit{e.g.,}~\cite{canetti2013obfuscating,brakerski2014black,brakerski2014virtual,barak2014protecting,bitansky2016indistinguishability,mahmoody2015more}) showed that an INDP obfuscator can be more powerful than merely providing INDP under idealized models.  

Besides, current graded encoding schemes can be considered as secure but still need to be carefully explored.  Both the GGH and CLT schemes are based on a new Graded Decisional Diffie-Hellman (GGDH) hardness assumption for multilinear maps.  The community generally agrees that the security of GGDH should be further explored, because it cannot be reduced to other well-established hardness assumptions, such as and NTRU for encryption over lattices~\cite{hoffstein1998ntru}.  Indeed, there are several investigations on cryptanalysis (\textit{e.g.,}~\cite{hu2016cryptanalysis,minaud2015cryptanalysis,cheon2016cryptanalysis,coron2016cryptanalysis}) or proposing newly patched schemes (\textit{e.g.,}~\cite{barak2014protecting,boneh2014immunizing,garg2016obfuscation,badrinarayanan2016post}).  However, no severe security flaw has been founded so far that would obsolesce the approach.

\subsection{With new Evaluation Properties}
There are two investigations (\textit{i.e.,}~\cite{dalla2005semantic,kuzurin2007concept}) which coincide with our results.  Kuzurin~\textit{et~al.}~\cite{kuzurin2007concept} observed that there are considerable gaps between practical and theoretical obfuscation.  On one hand, the security properties in theoretical or model-oriented obfuscation are too strong; on the other hand, we have no formal security evaluation approach for practical or code-oriented obfuscation.  They proposed to design specific security properties for particular application scenarios, such as constant hiding and predict obfuscation.  Preda and Giacobazzi\textit{et~al.}~\cite{dalla2005semantic} found that existing obfuscation evaluation metrics are textual or syntactic, which ignore the semantics.  They propose to employ a semantic-based approach to evaluate the potency of obfuscation.  To this end, they employ abstract interpretations to model the syntactic transformation of obfuscation with semantic-based approach~\cite{cousot2002systematic}.  A semantic-based obfuscation transformation is defined as $\tau[\mathbb{P}]$.  The obfuscation $\tau$ is potent if there is a property $\alpha$ such that $\alpha(S[\mathbb{P}]) \neq \alpha(S[\tau[\mathbb{P}]])$.  Such properties can be as simple as the sign ($\{+,-,0\}$) of a variable or a complex watermarking~\cite{cousot2004abstract}.  Moreover, the authors have conducted several preliminary investigations (\textit{e.g.,}~\cite{dalla2005control,dalla2006opaque,dalla2007code,giacobazzi2008hiding}) on employing the ideas to obfuscate simple programs.  

Note that all the investigations are still very preliminary.  There is still a large room for improvement in secure and usable obfuscation.

\section{Conclusions}\label{sec:conclusion}
To conclude, this work explores secure and usable program obfuscation in the literature.  We have surveyed both existing code-oriented obfuscation approaches and model-oriented obfuscation approaches, which exhibit gaps and connections in between.  Our primary result is that we do not have secure and usable program obfuscation approaches, and the main reason is we lack appropriate evaluation metrics considering both security and usability.  Firstly, we have no adequate security metrics to evaluate code-oriented obfuscation approaches.   Secondly, the performance requirement for model-oriented obfuscation approaches is too weak, and the security requirements might be too strong.  Moreover, we do not know how to apply model-oriented approaches to obfuscating general codes.  Our survey and result would urge the communities to rethink the notion of security and usable program obfuscation and facilitate the development of such obfuscation approaches in the future.

\bibliographystyle{IEEEtran}
\bibliography{survey}


\end{document}